\documentclass[aps,pre,onecolumn,superscriptaddress,nofootinbib]{revtex4-2}

\usepackage{graphicx}   
\usepackage{dcolumn}   
\usepackage{bm}              

\usepackage{graphicx}
\usepackage{dcolumn}
\usepackage{bm}

\usepackage{prettyref}
\usepackage{xcolor}
\usepackage{ulem}
\usepackage{setspace}

\usepackage[left=2cm,right=2cm,top=2cm,bottom=2cm]{geometry}

\usepackage{amsmath, amssymb, amsfonts}
\usepackage{amsmath}
\usepackage{amssymb}
\usepackage{mathtools}
\usepackage{bm}
\usepackage{physics}

\usepackage{graphicx}
\usepackage{float}
\usepackage{hyperref}
\usepackage[nameinlink,capitalize]{cleveref}

\crefname{equation}{Eq.}{Eqs.}
\crefname{figure}{Fig.}{Figs.}
\crefname{table}{Tab.}{Tabs.}
\crefname{subsection}{Sec.}{Secs.}
\crefname{appendix}{App.}{Apps.}

\usepackage{xstring}

\usepackage{lineno} 

\begin{document}


\parindent0mm
\setstretch{1.0}

\title{Comment on phase vortex lattices in neutron interferometry }

\author{Wolfgang Treimer}
\affiliation{Department of Mathematics,  Physics and Chemistry Berliner Hochschule fuer Technik, D-13353 Berlin, Germany}
%
\author{Terrence Jach}
\affiliation{Material Measurement Laboratory, National Institute of Standards and Technology, 100 Bureau Drive, Gaithersburg, Maryland 20899, USA}
\author{Martin Suda}
\affiliation{{\color{black}Optical Quantum Technologies}, Center for Digital Safety and Security, AIT  Austrian Institute of Technology GmbH, A-1210 Vienna, Austria}
\author{Frank Haußer}
\affiliation{Department of Mathematics,  Physics and Chemistry Berliner Hochschule fuer Technik, D-13353 Berlin, Germany}


%
\setstretch{1.2}
\begin{abstract}
\vspace{0.25cm}
\parindent0mm

We present a phase-based  theoretical description of the four-plate crystal neutron interferometer (LLLL geometry) originally introduced by Heinrich \textit{et al.} [Z.\ Phys.\ B \textbf{72}, 357 (1988)].  The LLLL   interferometer  can be interpreted as two coupled LLL interferometric loops that share partial beam paths, leading to characteristic interference phenomena beyond the standard three-plate configuration. 
Using a coherent beam-splitting model we derive the forward intensity in terms of loop phase sums and recover the cosine-sum structure reported in Refs.~\cite{Heinrich1988,Geerits2023}.
 
Based on these results we analyze  the phase dependence of the forward beam in the presence of localized phase objects, such as wedges inserted into individual partial beams. 
In experiments presented in \cite{Geerits2023}, the LLLL interferometer is used to generate a periodic array of intensity in two dimensions by which the authors claim to demonstrate a   phase vortex grating of significant extrinsic  orbital angular momentum (OAM).
However, the same interference patterns are predicted by
simulating a coherent superposition of  {beam paths} along different interferometer loops and experiencing phase shifts in each loop due to two wedges that are perpendicular to each other, without assuming any neutron OAM.
\end{abstract}

\maketitle

%
%
\setlength\parindent{0pt}
\date{\today}

\section{Introduction}

{Orbital angular momentum of quantum particles is a topic that has received considerable attention for some time
\cite{Beth1936,Allen1992,Barnett1994,Soskin1997,Mair2001,Yao2011,
Bliokh2017,Krenn2017,Fickler2018}.
 In its most meaningful interpretation, it consists of the modification or creation of a wave packet or wave function to carry additional angular momentum quite distinct from the intrinsic angular momentum attributed to spin. After its initial prediction [\cite{Beth1936,Allen1992}] and demonstration \cite{Mair2001,Yao2011,
Bliokh2017,Krenn2017,Fickler2018}  with photons in laser beams, it has been convincingly demonstrated with electrons (e.g.\cite{Bliokh2017}).\\ 

Wave functions that have been modified to carry orbital angular momentum have also been described as “vortex states.” This term, however, has been used to describe situations where quantum particles do not carry additional individual degrees of angular momentum and is subject to confusion.\\

Encouraged by the case of electrons, several investigators have attempted to create orbital angular momentum states in neutrons \cite{Clark2015,Sarenac2016,Sarenac2022,Sarenac2024}. This has led to the publication of results which we feel are misleading. The main consideration in the creation of intrinsic neutron orbital angular momentum states (n-OAM), in distinction to photons, is their fermionic nature. The creation of wave packets that individually carry OAM requires coherent phase shifting that is inherently not translationally symmetric. This constrains the process to the device scale and dimensions of an actual wave packet. For laser light, which is bosonic, a coherent state exists, such that the entire beam may be considered as the wave packet. This makes the helical phase modification of the state demonstrable on a scale of millimeters. It has also been demonstrated  that a sum of three beams created in an optical interferometer with slightly diverging wave vectors and precise phase shifts will create a spatial lattice, labeled “vortex states,” which are described by alternating regions of positive and negative orbital angular momentum \cite{Kurzynowski2006,Kurzynowski2010,Kurzynowski2012}. Again, because of their bosonic nature, isolated photons created in this manner would demonstrate inherent orbital angular momentum, in addition to that contributed by their spin.\\

Unfortunately, the demonstration of apparent OAM by  neutron interferometry is inherently insufficient. It has even been demonstrated that laser photons, whose transverse coherence is reduced to the point where the dimensions of a wave packet are no longer sufficient to undergo coherent phase shifting by a macroscopic phase plate, still show an interference pattern identical to what would be expected from photons genuinely in an OAM state \cite{Cappelletti2020}.\\ 

{A neutron wavepacket possesses intrinsic orbital angular momentum if its quantum state is an eigenstate (or a well-defined superposition of eigenstates) of the orbital angular momentum operator with respect to its own centroid and exhibits global transverse coherence.
Thus, a  neutron wavefunction $ \psi_{\mathrm{out}}(r,\phi)
=  e^{iq\phi}\psi_{\mathrm{in}}( r,\phi)  $  resembles an OAM eigenstate.
However, Cappelletti and Vinson  showed experimentally   that this is only valid if the neutron transverse coherence length $\xi_\perp $ is comparable to or larger than the radius of the spiral phase structure intended to create it,  $ \xi_\perp \gtrsim R_{\mathrm{SPP}} $ \cite{Cappelletti2020}.\\

If instead $ \xi_\perp \ll R_{\mathrm{SPP}} $, the neutron only probes a local phase gradient, and the outgoing state cannot be described as a globally coherent vortex state.}
}
To be clear, wave packets of small transverse coherence individually sample small regions of any phase shifting device and undergo a scalar phase shift that creates an identical positive or negative reinforcement in an interference situation. This can create apparent “vortex states” which are not vortex states at all.\\

In the case of neutrons, their fermionic character and typically microscopic transverse coherence make it impossible to create n-OAM states in any macroscopic manner. 
Therefore,  it has been demonstrated that several results attributed to the creation of n-OAM states \cite{Clark2015,Sarenac2016,Sarenac2022,Sarenac2024} can be easily reproduced by the transmission of ordinary neutrons through the same sequence of devices \cite{Cappelletti2018,Cappelletti2020,Jach2022,Treimer2023,Treimer2024,Treimer2025}.\\

Because the elastic scattering of ordinary neutrons in matter leads to scalar phase shifts easily associated with an index of refraction, neutrons are refracted by a wedge, or even by a wedge cylindrically wrapped around an axis \cite{Clark2015}. The refraction from such a device leads to a deflection of individual neutrons in an apparently spiral character, even in a device that is macroscopic relative to the transverse dimensions of a neutron wave packet. 
For a spiral wedge whose discontinuity is equal to the thickness that gives a scalar phase shift of  $ 2\pi $,  or integer multiples, a dimensional analysis attributes each neutron as conveying, through { $\vec{r} \times \vec{p} = -i\hbar (\vec{r} \times \nabla) $ }, an angular momentum multiple of  $\hbar  $ to the macroscopic device \cite{Clark2015}.\\

This detail, which is observable with ordinary neutrons and a macroscopic device of any dimension scale, has given rise to the notion that there are both “intrinsic OAM” and “extrinsic OAM”, where the former are genuine quantum states of individual particles, and the latter are merely array patterns of neutrons whose ensemble of transverse wavevectors locally demonstrates a curl or chirality.
We assert that such patterns, while perhaps interesting, have no quantum character, and the name “extrinsic OAM” is a misnomer. Results that have clearly fallen into the latter category have been identified as somehow connected with entangled states, informational qbits, or other genuinely quantum applications.\\

We comment here on a recent publication\cite{Geerits2023} which used the method of Kurzynowski  \cite{Kurzynowski2006,Kurzynowski2010,Kurzynowski2012}  using wedges to create three beams in a neutron interferometer resulting in a lattice of “extrinsic OAM”. Interferometers for neutrons function by means of Laue diffraction. 
It has been demonstrated that the wave fields in Laue interferometers are markedly different from those of simple optical interferometry \cite{RauchTreimer1974,Treimer1975,Treimer1978,Treimer1978a,Suda2006,RauchWerner2015}.
 We describe the  mechanism of a Laue interferometer and demonstrate that the observed patterns are simply the result of ordinary neutrons in an interferometer undergoing additional  scalar phase shifts that distort the uniform pattern of maxima and minima into a two dimensional lattice of enhanced intensity.

\section{Perfect crystal interferometry}

Perfect-crystal neutron interferometry provides a unique platform for precision studies of quantum coherence, phase shifts, and multi-path interference of matter waves. 
The standard three-plate Laue interferometer (LLL geometry, \cite{RauchTreimer1974,Treimer1975, Treimer1978,Treimer1978a}) has been extensively investigated and is well understood in terms of two-path interference \cite{RauchWerner2015}.\\  

An extension of the LLL type interferometer is the four-plate crystal interferometer (LLLL geometry), first analyzed by Heinrich \textit{et al.}~\cite{Heinrich1988}. 
The additional Laue plate introduces additional coherent splitting and recombination stages, and allows for manipulating three  independent interferometer  loops. 
Geerits \textit{et al.}~\cite{Geerits2023} used this LLLL neutron interferometer, claiming to  generate a phase vortex grating with significant extrinsic orbital angular momentum (OAM), $ <L_z> \approx 0.35 $, on a length scale of ca. 220 $ \mu $m transverse to the direction of propagation.\\

We recall several publications in which a neutron interferometer was used to detect the orbital angular momentum of neutrons and to demonstrate holography involving orbital angular momentum (see, for example, \cite{Clark2015,Sarenac2016}). 
Shortly thereafter, however, critical articles appeared that questioned whether the observed phenomenon could indeed be attributed to the angular momentum of neutrons (see,  \cite{Cappelletti2018,Cappelletti2020, Jach2022}). Further articles showed that in the mentioned publications, the orbital angular momentum could not have been detected at all, since the lateral coherence lengths were too small and too variable \cite{Treimer2023,Treimer2024,Treimer2025}.\\

The moiré effect caused by the wedges and twisted crystal plates of an LLL interferometer has been known, due to the work of Bonse and M. Hart, since 1965 \cite{Bonse1965}.
The neutron interferometer described in \cite{RauchTreimer1974,Treimer1975} was first studied in detail at U. Bonse's institute (University Dortmund)  using X-rays before being successfully tested with neutrons at the Atomic Institute in Vienna.
All the expected interference effects due to the wedges and rotations of the crystal plates could be perfectly explained as moiré effects.  \cite{Treimer1975,Treimer1978, Treimer1978a}.\\

This paper provides a concise theoretical description of the LLLL interferometer, with a focus on phase modulation caused by localized phase objects such as wedges.
We formulate the problem in terms of beam path phases and use a coherent beam-splitting model, excluding neutron orbital angular momentum (OAM). Finally, we compare the resulting expressions with recent work by Geerits \textit{et al.}~\cite{Geerits2023}.\\  

\begin{figure}[hbtp]
\centering
\includegraphics[width=12cm]{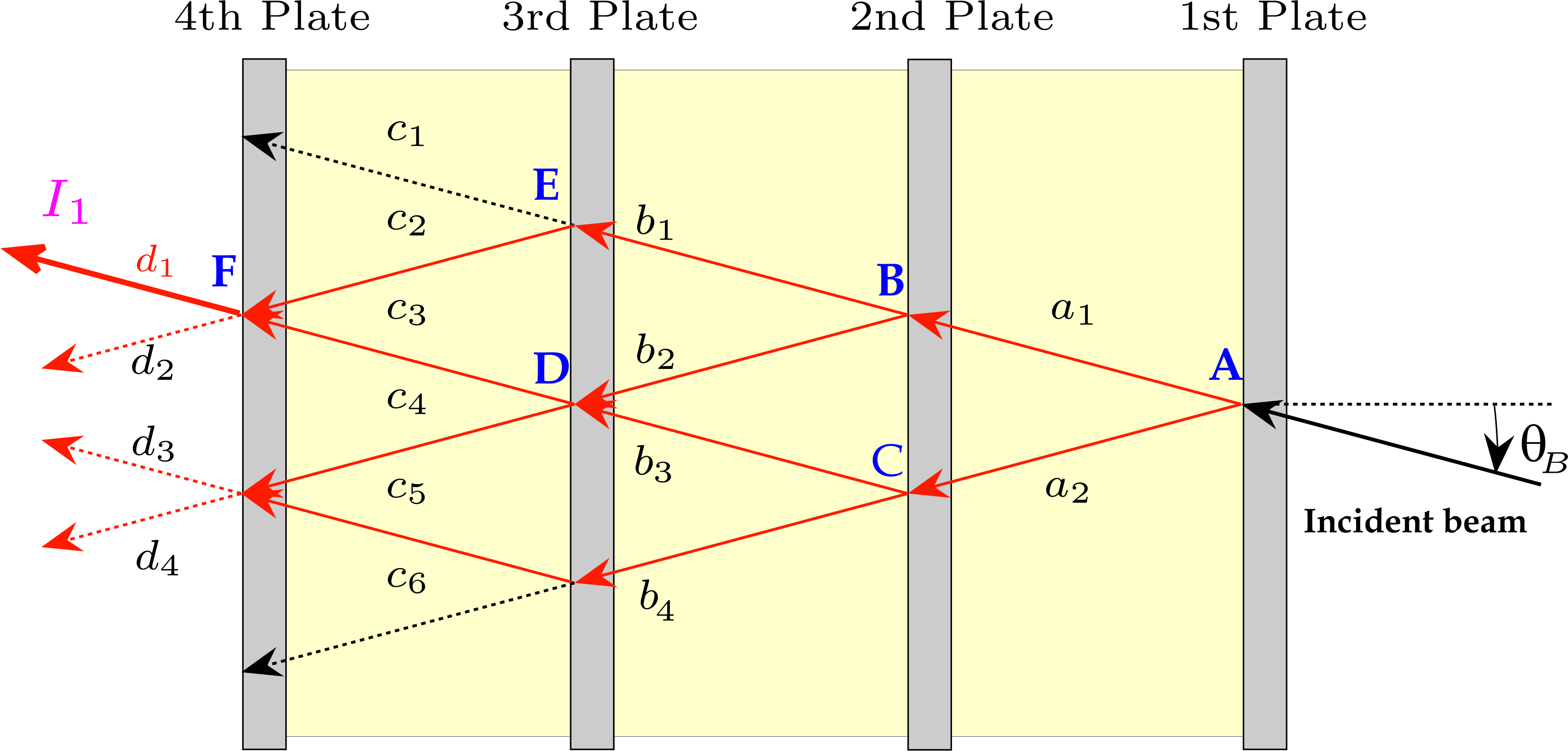}
\caption{Beam splitting by Laue diffraction at A, B, C, D, E and F. Plates 1 .. 4 are perfect crystals of equal thickness and the distance between the crystal plates is the same. $ \theta_{B} $ is the Bragg angle.}
\label{fig:4PlateIFM_I}
\end{figure}

We consider the four-plate perfect-crystal interferometer in Laue geometry as described in\ Ref.~\cite{Heinrich1988}.  The first plate coherently splits the incident wave into two partial beams $a_1$ and $a_2$.  At the second plate each partial beam undergoes again Laue diffraction, generating four coherent beams $b_1,b_2,b_3,b_4$. At the third plate further splitting produces beams $c_i$, of which the exit beams   $c_1$ and $c_2$ at E and  $c_3$ and $c_4$ at  $D$ {are considered and} where only  $c_2$ and $c_3$   are of primary interest  (\cref{fig:4PlateIFM_I}).\\

Rays  $a_1 $,  $a_2  $, $b_2 $  and $b_3 $ are  the transmitted and  reflected  ones  from beam splitting at A, B and C {and} superimpose at D, creating coherent rays $c_3  $ and $c_4 $. Thus, loop I  ($\widehat{ABDCA} $, interferometer I)  corresponds to a classical LLL interferometer. 
Loop II ($\widehat{BEFDB} $)  also looks like a LLL interferometer, but the phases of the  rays $b_1$, $b_2 $,  $c_2 $  and $c_3 $ depend on phase shifting objects  in interferometer loop I.\\
 
A useful interpretation is obtained by decomposing the ray network into two coupled interferometric loops, which  are first considered separately (see \cref{fig:4PlateIFM_II}). 
The first loop is denoted $\widehat{ABDCA} $ and corresponds to an effective LLL interferometer producing the output beams $c_3$ and $c_4$.
A phase shifting object (wedge) in one of the sub-beams $ a_1,  a_2,  b_ 2,  b_3 $ modulates the exit beams at D,  $c_3$ and $c_4$, and it does not matter in which subbeam $ a_1 \ldots  b_3 $ the phase object is located.\\

The second loop, denoted $\widehat{BEFDB} $, shares partial beam segments with the first loop and modifies the phase relations at the output rays d$ _{1} $ an d$ _{2} $.  
The coupled-loop character is reflected in the appearance of interference terms depending on both loop phases simultaneously. 
It is interesting to note that, if interferometer loop I delivers  perfect diffraction, interferometer loop II would not function as an interferometer if the partial wave {traveling} along path $\widehat{ABC} $ {were} shifted by $\pi $ relative to the partial wave along path $\widehat{ACD} ${,} because then  the amplitude in $ c_3  = 0 $.  The intensity $ I_{c_3} $ at D is given in the forward direction ($ c_3 $ - ray) as $ I_{c_3} \propto [  1+\cos\!\big(\phi_1-\phi_2\big) ] $,  if $ \phi_1 $ is the sum of phase shifts in rays   $ a_1 + b_2 $ and $ \phi_2  $  the sum of phase shifts  in rays $ a_2  +  b_3 $. 
For   $  \phi_1-\phi_2  = \pi  $   one gets  $ I_{c_3} $ =  $  1+\cos\!\big(\pi\big)  $   and  from particle conservation  $ I_{c_4} = 1 - I_{c_3}  =  1$ {. That is to say}, there is no intensity in {the $ c_3 $ - beam}, so no interference  at F is possible.
A phase-shifting object in a beam whose intensity is zero due to destructive interference does not contribute to the interference, as already mentioned in \cite{Heinrich1988}.
Therefore phase plate $PS_1  $ (\cref{fig:4PlateIFM_II})  controls the  intensity distribution  at D  and {the contribution of beam} $ c_3 $ to interference. \\

The intensity of beam $ c_3 $ is therefore  determined by phase flag $ PS_1  $  and by  a phase-shifting object in {interferometer} loop I. 
Note that the position at which the wedge $ W_1 $ is placed in loop I determines which interference pattern is observed at F. Different patterns are obtained depending on whether $ W_1 $ is located in beam $ a_2 $ or $ b_3 $, or whether the wedge $ W_1 $ is located in beam $ a_1 $ or $ b_2 $. The intensity $I_1$ is of particular interest because all three partial rays $(a_1,b_1,c_2,d_1),(a_1,b_2,c_3,d_1),(a_2,b_3,c_3,d_1)$ are transmitted twice and reflected twice, and therefore have the same amplitude and phase in the empty interferometer ({This is true even after a full consideration of all branches in Laue dynamical diffraction}).


\section{Calculation of forward intensity $ I_{1} $ }

\begin{figure}[h]
\centering
\includegraphics[width=12cm]{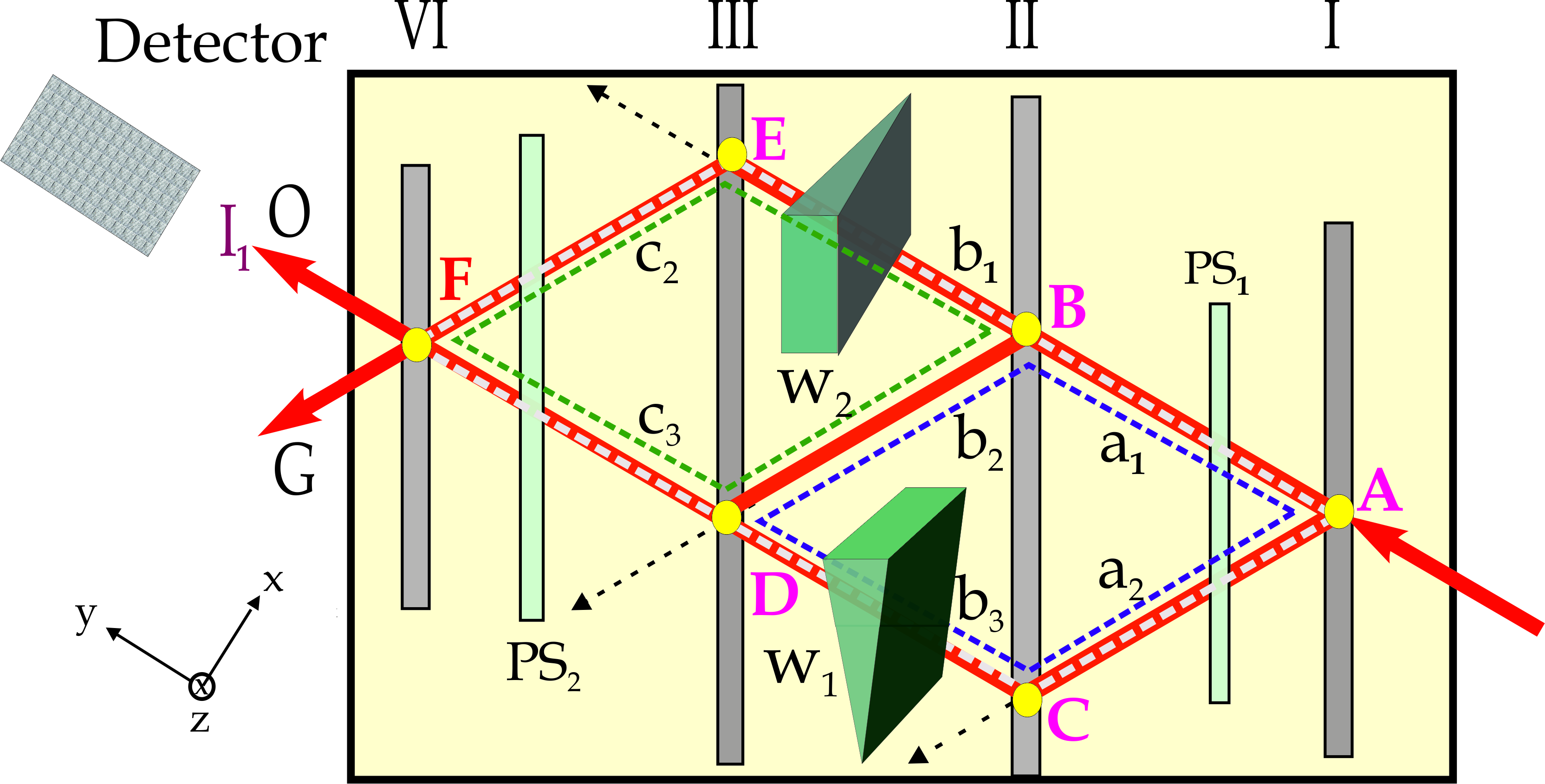}
\caption{Layout of a four plate crystal interferometer, I ... IV denote the four 	 crystal plates, O an G  are the  beams behind the interferometer, only the O 	  beam is discussed, 
PS$_1$ and PS$_2$ are phase-shifting plane plates that can  optionally shift the phase and change the intensity of loops.
{Between plate II and III the neutron beam is parallel to $ b_1 $  and $ b_3 $ , respectively, in the y-direction and incident perpendicularly to the respective wedge surface of $  W_1 $  and  $ W_2 $.}
}
\label{fig:4PlateIFM_II}
\end{figure}

Thus, successive Laue - splitting  at the four perfect crystal plates I ... IV at points A, B, C, D , E and F  creates two coupled {interferometers}, loop I and  loop II, {whose beams}  $ a_{1,2},  b_{1 \cdots 3} $ and $ c_{2,3} $ superimpose  and contribute to interference in F (see  \cref{fig:4PlateIFM_II}).\\

To illustrate  coherent beam splitting and superpositions, we will for simplicity's sake first assume perfect, thick single crystals that produce a 50\:\%  coherent Laue beam splitting ({at the} ideal Bragg condition) at points A, B, C, D, E and F, all with the {assumption of identical} amplitude. 
{Furthermore,  only the  wave paths leading to $I_1$ are considered.\\
Let $\psi_0 $ be the incident plane wave at point $A$.  After the first splitting at A{,} one has partial rays 
\[ a_1 = \frac{1}{2}\psi_0 e^{i\alpha_1}, 
\quad
a_2 = \frac{1}{2}\psi_0 e^{i\alpha_2}.
\]
{Here,} $ \alpha_{1,2} $ are possible phase shifts if objects were inserted between $ \widehat{AB} $ and/or $ \widehat{AC} $, or if PS$_1$ is not parallel to plate I.
At the second plate, at point B, $a_1$ splits into $b_1$ and $b_2$, while at point C, $a_2$ generates $b_3$  and  $b_4$. Possible phase shifts, $ b_{1 \cdots 3 } $, ($b_4 $ will be neglected)  are 
\[	b_1 = \frac{1}{2}a_1 e^{i\beta_1}
		= \frac{1}{4}\psi_0 e^{i(\alpha_1+\beta_1)}, \quad 	b_2 = \frac{1}{2}a_1 e^{i\beta_2} 
		= \frac{1}{4}\psi_0 e^{i(\alpha_1+\beta_2)}, \quad 
		b_3 = \frac{1}{2}a_2 e^{i\beta_3} = 
		\frac{1}{4}\psi_0 e^{i(\alpha_2+\beta_3)}   
\]
The relevant output amplitudes at E and D are
\[
	c_2 = \frac{1}{2}b_1 = \frac{1}{8}\psi_0 e^{i(\alpha_1+\beta_1)}, \quad 
	c_3 = \frac{1}{2}b_2 + \frac{1}{2}b_3
	     = \frac{1}{8}\psi_0 e^{i(\alpha_1+\beta_2)}
	     + \frac{1}{8}\psi_0 e^{i(\alpha_2+\beta_3)}.
\]
The output amplitudes before F are
\[
c_2 = \frac{1}{2}b_1e^{i\gamma_2} = \frac{1}{8}\psi_0 e^{i(\alpha_1+\beta_1+\gamma_2)}, \quad 
c_3 =  \left(\frac{1}{2}b_2 + \frac{1}{2}b_3 \right)e^{i\gamma_3}
	     = \frac{1}{8}\psi_0 e^{i(\alpha_1+\beta_2+\gamma_3)}
	     + \frac{1}{8}\psi_0 e^{i(\alpha_2+\beta_3+\gamma_3)}
\]
%
%
%
The final amplitude $ i_1 $ in {the} forward direction is the coherent sum of $ c_2 $ and $ c_3 $, 
\[
i_1 = \frac{1}{16}\psi_0 \left( e^{i(\alpha_1+\beta_1+\gamma_2)} + e^{i(\alpha_1+\beta_2+\gamma_3)} + e^{i(\alpha_2+\beta_3+\gamma_3)} \right)
\] 
and the intensity   $ \vert i_1\vert^{2} $ = $ I_1 $  can be calculated  in  the common case, omitting  phase objects in  rays $ c_{2,3} $ as:

		\begin{equation}
		I_1 
		= \mathrm{const}\cdot
		\left[3
		+2\cos(\beta_1-\beta_2)
		+2\cos\big((\alpha_1-\alpha_2)+(\beta_1-\beta_3)\big)
		+2\cos\big((\alpha_1-\alpha_2)+(\beta_2-\beta_3)\big)
		\right].
		\label{eq:I_general}
		\end{equation}
Here $\alpha_i$, $\beta_i$ are the phase shifts of the partial beams $a_i$, $b_i$. {A more detailed calculation can be found in \cite{Suda2006}.} For the special choice as assumed in \cite{Geerits2023} and in this  paper,  
no phase shifting objects are in rays  $ a_1,  a_2 $   and  $ b_2 $,    $ \alpha_1 =\alpha_2 = \beta_2 = 0 $,  so one obtains
		\begin{equation}
		I_1 = \mathrm{const}\cdot  \left[ 3 +2(\cos(\beta_1) +
		\cos(\beta_1-\beta_3) +\cos(\beta_3))\right],
		\label{eq:Isimplified}
		\end{equation}
$\beta_{1} $  and $\beta_{3} $ are three-dimensional, homogeneous phase-shifting objects (wedges) that can be placed in one of the LLLL neutron interferometer's rays, as shown in \cref{fig:4PlateIFM_II}.\\

In \textit{Heinrich et al 1988} (\cite{Heinrich1988}) the corresponding intensity   $ I_{0}(\chi_A,\chi_{B}) $  was written  as
		\begin{equation}
		I_{0}(\chi_A,\chi_{B}) 	= 
		K_2\left[3 +2\Big( \cos(\chi_A+\chi_{B})
		+\cos(\chi_A) 	+\cos(\chi_{B}) 	\Big) 	\right].
		\label{eq:I0_chi}
		\end{equation}
$ K_2  $ is a constant given in \cite{Heinrich1988},  and  with   $  {\alpha _1} = \beta _2 = \alpha _2 = 0,  
{\chi _A} =  - {\beta _3} $ and $ {\chi _B} = {\beta _1}  $
we get the identical equation  to \cref{eq:Isimplified} 
	\begin{equation}
	I_{0}(\beta_3,\beta_1) = K_2\left[ 3 +2\Big(\cos(\beta_1 - \beta_3)
	+\cos(\beta_3) +\cos(\beta_1) \Big) \right].
	\label{eq:I0_chi}
	\end{equation}
Now, let us turn to the paper by Geerits {\it{et al.}} (2023) \cite{Geerits2023} and its notation. \\

Since we intend to adopt the formulas from [2] as is, we first must point out the differences in the coordinate systems used in this paper and in  [2]. In this paper, in Fig. 2, we are using an (x,y,z) coordinate system, whereas Geerits {\it{et al.}} [2] uses an (y,z,x) coordinate system.
Therefore, the lateral coordinates are (x, z) in our figure, but (x, y) in \cite{Geerits2023}. This difference  appears in equations \cref{equ:Geerits_Glg.2} and \cref{eq:I_r0} as well as in equations \cref{equ:Geerits_Glg.2 u a} and \cref{equ:L_z} below, which were taken directly from \cite{Geerits2023}. 
In  \cite{Geerits2023}  one starts with 
\begin{equation}
\psi_1(\mathbf{k}) = \frac{1}{\sqrt{3}} \left[ \psi_0(\mathbf{k}) + e^{i\alpha_1} \psi_0(\mathbf{k}-\mathbf{k}_{\perp} \hat{y}) + e^{i\alpha_2} \psi_0(\mathbf{k}-\mathbf{k}_{\perp} \hat{x}) \right].
\label{equ:Geerits_Glg.2}
\end{equation}
%
We interpret  that   $\psi_0(\mathbf{k})$   corresponds to the unshifted reference beam, the terms  $ e^{i\alpha_1} \psi_0(\mathbf{k} - k_\perp \hat{y}) $ and    $ e^{i\alpha_2} \psi_0(\mathbf{k} - k_\perp \hat{x}) $ 
describe the beam shifted in the $y$-direction  and $x$-direction, respectively,  in momentum space, which are imprinted with controlled phases  $\alpha_1$  and  $\alpha_2$,   $\hat{x}$ and $\hat{y}$ are perpendicular to $\mathbf{k} $.  
Thus, in \cref{equ:Geerits_Glg.2}  the transverse phase shifts $ k_{\perp} $ are the phase shifts  of the prisms (wedge elements) in the beam paths $ b_3 $ and $ b_1 $ as  described here (see \cref{fig:4PlateIFM_II}). 
\cref{equ:Geerits_Glg.2} (Eq(2) in \cite{Geerits2023})  can be rewritten as Intensity as 
		\begin{equation}
		I(\mathbf{r}_0)=\frac{1}{3}\left[ 3 +2 \Big(\cos(k_\perp y+\alpha_1)
		+\cos(k_\perp x+\alpha_2) 	+\cos\bigl(k_\perp (x-y)
		+\Delta\alpha\bigr)\Big) 	\right].
		\label{eq:I_r0}
		\end{equation}
Comparing  Eq. (23) in \cite{Geerits2023},  $ \Delta\alpha  = \alpha_1 - \alpha_2 $,   $ \alpha_1 \text{  and  } \alpha_2  $ correspond simply to  phase  shifts    $ \alpha_1$  in ray  $a_1$,   and  $ \alpha_2  $  in  ray  $ a_2$ (see   \cref{fig:4PlateIFM_II}). 
{The equations for the exit  intensity $ I_1 $  (Eq (2)  this paper and $ I_{0}(\beta_3,\beta_1) $  Eq (4), given in [1]) agree perfectly. Eq(5) given in Geerits {\it{et al.}} looks different but coincides with both, Eq(2) and Eq(4). It can be shown that \cref{eq:I_r0} and \cref{eq:Isimplified} (both this paper) are equivalent descriptions of the interaction of neutrons with wedge-shaped phase shifters, with the difference that refraction, as mentioned in \cite{Geerits2023}, plays no role. Rather, a wedge with a small vertex angle of 5° merely causes a position-dependent phase shift.
}
Taking into account a possible intrinsic phase $ \phi_0 $ of the interferometer, phase shifts caused by $ PS_{1,2} $, and/or individual phase shifts $\Delta \phi_{i = 1 \cdots 3}$ in the crystal plates due to distortions, lattice fluctuations, or point defects, we would  obtain: 
\begin{equation}
\begin{aligned}
I_{1}  \sim     [ const + \cos \left( \phi_0 +  (\beta_1 + \Delta\phi_{1}) \right) 
+  \cos \left( \phi_0 +  (\beta_1 - \beta_3 + \Delta\phi_{2}) \right) 
+  \cos \left( \phi_0 +  (\beta_3 + \Delta\phi_{3}) \right) ]  
\end{aligned}  
\end{equation}
%
%


\section{Calculation of the interference patterns}

For calculations of interference patterns from {a} crystal  neutron interferometer  we allow for interference between rays  only in the scattering plane {\color{red}\cite{Radon1917}}, defined by the incident neutron wave vector \bm{$\vec{k}$}  and the reciprocal lattice vector \bm{$ \vec G $}, because, as already mentioned above, and pointed  out in  \cite{Jach2022,Treimer2023,Treimer2024,Treimer2025}, due to the extremely different transverse coherence lengths {$ \sigma_z \ll \sigma_x  $, $ \sigma_{x} \approx {\text{5}\:\mu}m , \sigma_{z} \approx  50 \: nm$, i.e. about two orders of magnitude,} there is  no interference  possible out of the scattering plane (see \cref{fig:interference pattern wedge}).\\
\begin{figure}[h]
\centering
\includegraphics[width=8cm]{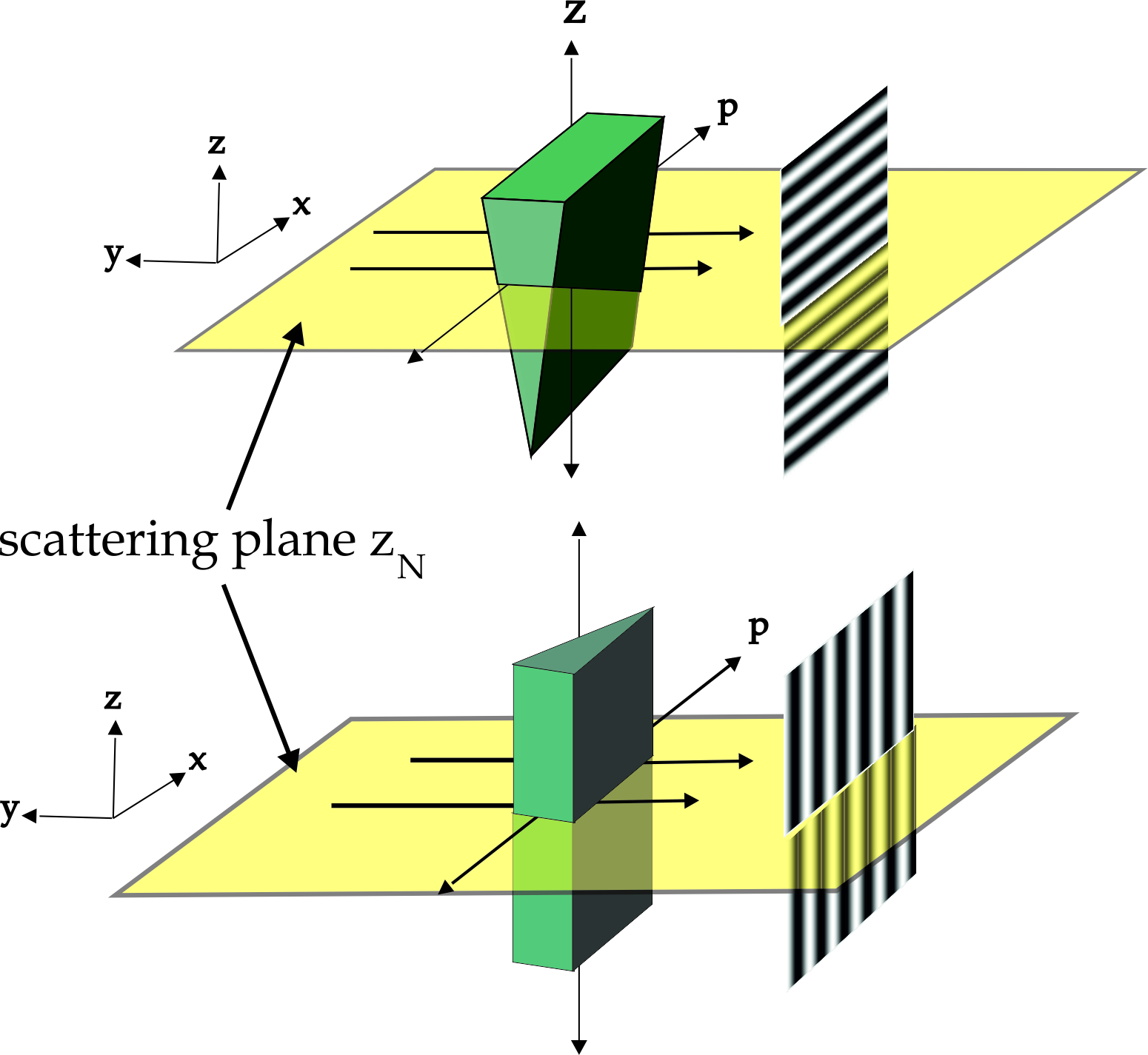} \\
\caption{Interference pattern of the vertical wedge  and  of  the  horizontal wedge. 
 For a vertical wedge, rays in the (y,z)  plane pass through the wedge at different thicknesses and therefore, they experience different phase shifts compared to rays in other (x,y) planes  In the  case of the horizontal wedge rays in the (x,y) plane traverse different wedge thicknesses and cause corresponding phase shifts (interference pattern see  \cref{fig:interference patterns of wedges}a,b).  Together, they form an interference pattern as shown in \cref{fig:interference patterns of wedges}f. }
 \label{fig:interference pattern wedge}
\end{figure}

Phase-shifting objects inserted into individual partial beams modify the relative phases of interfering amplitudes and thereby generate spatially structured interference patterns in the exit beams. 
A wedge as phase shifting object introduces a transverse phase gradient, resulting in fringe pattern  in the detector plane.  The orientation of the  pattern is directly connected with the orientation  of the phase shifting wedge in the neutron interferometer (see \cref{fig:interference patterns of wedges}a,b).\\

For a single wedge inserted into any of the partial beams belonging to loop $\widehat{ABDCA}$, the fringe pattern in the output $ c_{3,4} $ depends only on the accumulated phase difference between the two interfering paths  $ \{ a_1 , b_2 \} $   and $ \{ a_2 , b_3 \} $. 
Consequently, the fringe spacing and orientation are independent of the  location of a wedge within this loop. The complementarity of the two output beams in neutron interferometry signifies that maxima in $ c_3 $ correspond to minima in $ c_4 $. 
This can be detected if the 2D detector simultaneously detects $ c_4 $  with the detector in the O position.\\

If a second wedge is added, the resulting pattern depends strongly into which sub-beam it is positioned.  
Depending on its position, the second wedge can cancel out the phase gradient created by the first wedge. This suppresses the fringes. Alternatively, it can amplify the phase gradient, which increases the fringe density or produces an entirely different pattern. This would happen, for example, if the wedges were aligned perpendicular to each other. \\

{We calculated  the two-dimensional phase front of the aluminum wedges  $ W_1 $ and $ W_2 $ with parameter values given in \cite{Geerits2023}, i.e., neutron wavelength $\lambda_n =  0.192\, \mathrm{nm}$ and wedge apex angle $\alpha = 5^\circ$.  
This leads to a $\lambda$-thickness
\[
 D_{\lambda ,Al} =  \frac{2 \pi}{\lambda_n N_\mathrm{Al}bc_{Al}} = 157.4 \,\mu \mathrm{m}, 
 \]
 where $N$  denotes the number of atoms per unit volume and $bc$ the coherent scattering length of aluminum. 
 Inserting  $ W_{1}  \rightarrow b_1 $ and/or $ W_{2} \rightarrow b_3 $ leads to
 corresponding phase shifts $\beta_1$ and $\beta_3$ (with respect to some reference point $(x_0, z_0))$ as
 \begin{align*}
 \beta_1 (x, z) &= 2 \pi z \tan{\alpha}/D_{\lambda ,Al} , \quad \beta_3 (x, z) =  -2 \pi x \tan{\alpha}/D_{\lambda ,Al}.
 \end{align*}
The formula 
    \[	I_1 = \mathrm{const}\cdot  \left[ 3 +2(\cos(\beta_1) +
    	\cos(\beta_1-\beta_3) +\cos(\beta_3))\right],
    \] 
 (see \cref{eq:Isimplified} )
leads to the interference pattern presented in \cref{fig:interference patterns of wedges}(e).\footnote{The wave length of the inference pattern caused by a single wedge is $\tan{\alpha}/D_{\lambda ,Al} = 1.799\, \mathrm{mm}$}}
In  \cite{Heinrich1988} the measured  contrast of  the individual interferometer loops is 58\:\% for loop I and 47\:\% for loop II, corresponding to loop $\widehat{ABDC} $  and loop  $\widehat{BEFD} $. This was taken into account calculating \cref{fig:interference patterns of wedges}(f).
These phase-modulated partial waves from different propagation paths interfere before or within the crystal at points D and F and  together with the excited wave fields of the reflecting crystal lattices one measures a moiré pattern  \cite{Bonse1969,HartBonse1970,Deslattes1971}. \\

  \begin{figure}[h]
    \centering
    \includegraphics[width=12cm]{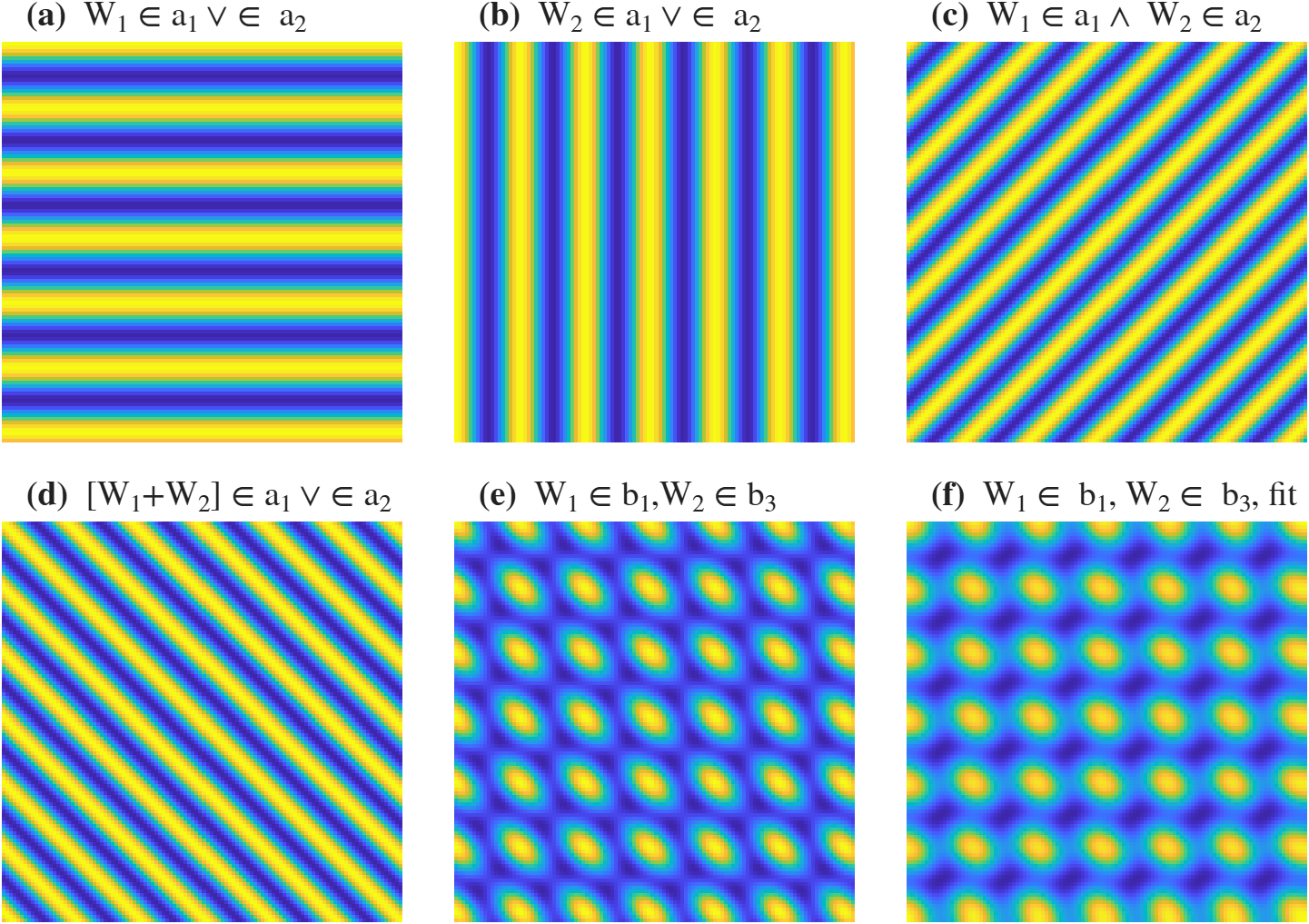} 
    \caption{Calculated interference patterns of one and two wedges in different paths of the neutron interferometer,  pattern (f) is the smoothed version of pattern (e), see position of wedges in \cref{fig:4PlateIFM_II}, (f) Interference patterns, taking into account the reduced contrasts of loops I and II at 47\:$\%$ and 58\:$\%$  (see \cite{Heinrich1988}). 
    Fig.5 shows a cropped version of Fig. 4(f), which allows us to compare our calculated interference pattern with that measured and fitted by Geerits {\it{et al.}}
} 
    \label{fig:interference patterns of wedges}
  \end{figure}

In \cref{fig:interference patterns of wedges} we show four interference patterns (a) – (d) in which the wedges $ W_1 $ and $ W_2 $ have been placed along different paths ($ a_1, a_2 $), to illustrate the effect of the wedges’ positions in the interferometer on the interference patterns.
 In the case of interference \cref{fig:interference patterns of wedges}(c) $ W_1 $  and $ W_2 $  were placed in different paths {($ \widehat{ABD} $ and $ \widehat{ACD} $)},  in  \cref{fig:interference patterns of wedges}(d)  $ W_1 $  and $ W_2 $   were in the same path, in ($ \widehat{ABD} $ or  in $ \widehat{ACD}) $. 
\cref{fig:interference patterns of wedges}(f)  shows the interference pattern if one wedge is in path  $\widehat {BE} = b_1$ and the other wedge placed in $ \widehat {CD} = b_3  $. 
 {This configuration is identical to what is done in \cite{Geerits2023} }.
The pattern in  \cref{fig:interference patterns of wedges}(f)  was calculated with the reduced contrast of the loops I ($  47\%  $) and loop II ($ 58\% $), and smoothed, respectively. \\

Let us consider  Eq.(2) in \cite{Geerits2023} (this paper  \cref{equ:Geerits_Glg.2}):  The test function $ \psi_t(\mathbf{r}) $ in \cite{Geerits2023} is constructed as  the superposition of two orthogonal linear momentum components:
\begin{equation}
\begin{aligned}
\psi_t(\mathbf{k})& = \frac{1}{\sqrt{3}} \left[ \psi_0(\mathbf{k}) + e^{i\alpha_1} \psi_0(\mathbf{k}-\mathbf{k}_{\perp} \hat{y}) + e^{i\alpha_2} \psi_0(\mathbf{k}-\mathbf{k}_{\perp} \hat{x}) \right]   \\
& \text{  rewritten    in  cylindrical coordinates (Eq. (4) in \cite{Geerits2023}) } \\
\Psi_t(\rho, \phi)& = \frac{1}{\sqrt{2}} \psi_0(\rho) \left[ \exp\left( i k_{\perp} \rho \sin\phi \right) + \exp\left( i \Delta\alpha \right) \exp\left( i k_{\perp} \rho \cos\phi \right) \right]
\end{aligned}
\label{equ:Geerits_Glg.2 u a}
\end{equation}
where  $k_{\perp}$ denotes the magnitude of the transverse wavevector, dictating the spatial frequency of the interference fringes,   $\phi = \arctan2(y, x)$ is the azimuthal angle and 
 $\Delta\alpha$ is the relative phase shift between the two orthogonal components (was set in \cite{Geerits2023} to $\pi/2$  'to form a phase vortex lattice'). 
 { $\Delta\alpha$ corresponds to the phase shift between the rays $ a_1 $ and $ a_2 $ (see \cref{fig:4PlateIFM_II}), i.e., a rotation of the phase flag PS1. The effect of $ \Delta\alpha $ is shown below (\cref{fig:flag rotation}).   
Thus, the ansatz of a test function is not obvious, and without the appropriate prior knowledge, one would neither suspect nor recognize a vortex lattice or an extrinsic n-OAM.
}

Eqs.(13,14)  given in \cite{Geerits2023}  use their  Eq.(8, 14)  for writing  the expectation value of the OAM operator $ L_z $  and   \(\langle L_z \rangle\)  as function of  \(k_{\perp }\) and  phase \(\Delta\alpha\), 
{
\begin{equation}
\langle L_z \rangle = \sin(\Delta\alpha)  \frac{ k_{\perp}^2 \sigma^2}{4N}  e^{-\frac{k_{\perp}^2 \sigma^2}{4}} =   
 \sin(\Delta\alpha)  \frac{ k_{\perp}^{2}  ( \sigma^2_x +\sigma^2_y)}{8N}  e^{-\frac{k_{\perp}^2 ( \sigma^2_x +\sigma^2_y)}{8}}    
 \label{equ:L_z}
\end{equation}
where $ \sigma  $ is the transverse lateral coherence length.
{Note the different coordinate systems used in Geerits \textit{et al.} and in this paper. }  In \cite{Geerits2023}  $ \sigma^{2}  $ is  defined as $  \sigma^2_x +\sigma^2_y $, assuming    $ \sigma^2_x  \sim \sigma^2_y $  \textit{which is a wrong assumption}, because of  $ \sigma_x\ll \sigma_y $ and  thus $ \sigma^{2} \approx \sigma_y^{2} $ (see  \cite{RauchWerner2015,Pushin2008}).  
Then $ \langle L_z \rangle $   becomes correctly
}
\[
\langle L_z \rangle = \sin(\Delta\alpha)\,
\frac{k_{\perp}^{2}\sigma_y^2}{8N}\,
\exp\!\left(-\frac{k_{\perp}^{2}\sigma_y^2}{8}\right).
\] \\
{($\sigma_x$} in \cite{Geerits2023} corresponds to $\sigma_z$ here.) 
 Therefore, $ \langle L_z \rangle  $ becomes "one-dimensional" {which} explains {why} the measured interference pattern in \cite{Geerits2023} looks like \cref{fig:interference patterns of wedges}f in this paper.  Interference occurs solely in the scattering plane; {vertical variations originate purely from phase shifts modulated perpendicularly to the scattering plane.} Thus, an OAM cannot be detected in this way.
\begin{figure}[hbtp]
\centering
\includegraphics[width=4.5cm]{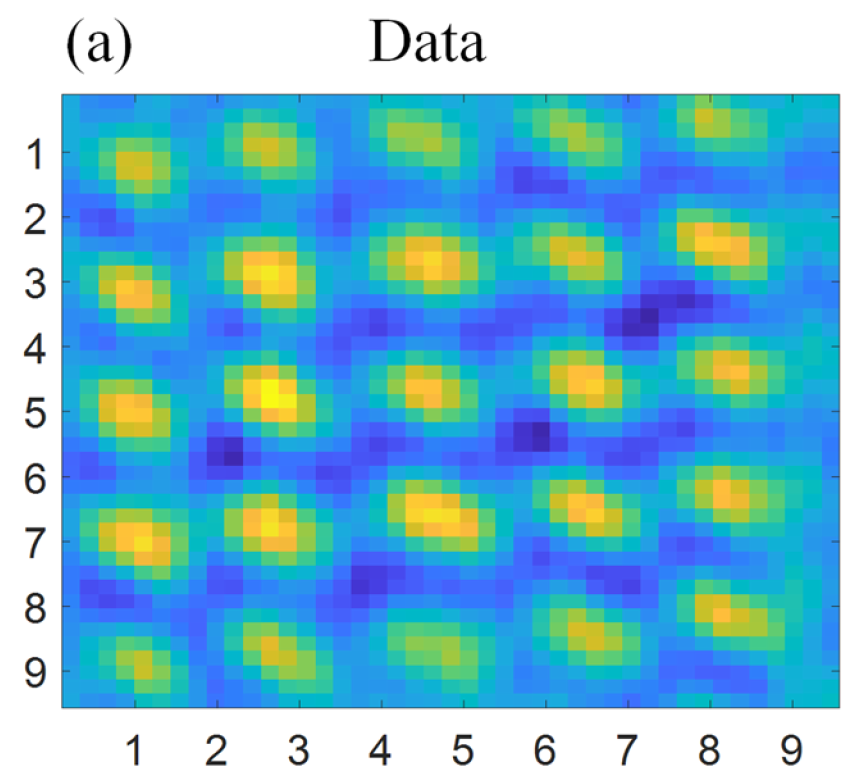} 
\includegraphics[width=3.9cm]{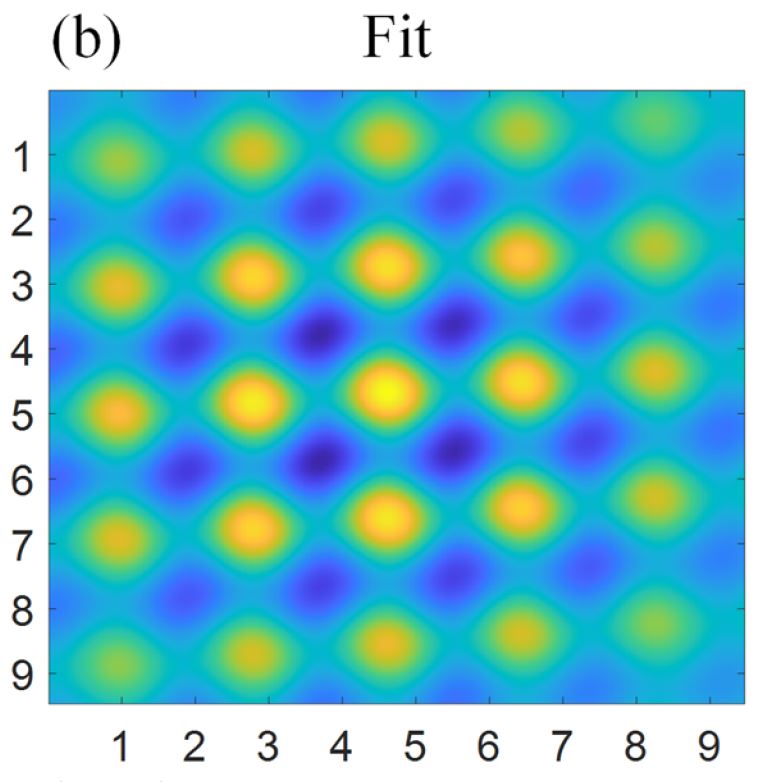} 
\includegraphics[width=3.5cm]{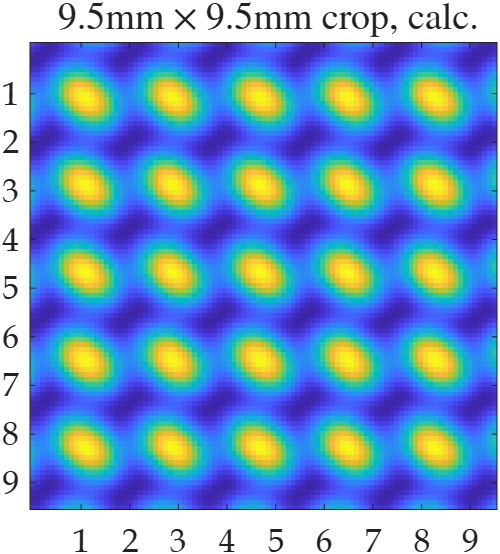} 
\caption{Comparison  of interference patterns.
Note that the bright structures in the image on the left are oriented diagonally towards the top left ( (a) Data ), a detail that was lost after the data was fitted (middle image, (b)  Fit), but which is clearly visible in the processed image on the right.
}
\label{fig:Comparison Figures 4b_ 5a}
\end{figure}

\cref{fig:Comparison Figures 4b_ 5a} (images (a) and (b)) shows the experimental result and data fit of the interference pattern with two wedges in the LLLL neutron interferometer, as depicted in {Fig. 4a and b in  \cite{Geerits2023} }, and our calculation (right image) using the same experimental parameters.\\


In \cref{fig:flag rotation} we demonstrate the change of the interference pattern $ I_1 $, if the phase flag $ PS_1 $  (see \cref{fig:4PlateIFM_II})  is rotated from $ 0^{\circ} $ to $ 6^{\circ} $ . 
A  rotation $ PS_1 $  creates a phase shift $ \Delta\alpha = \alpha_1  - \alpha_2 $ (the same $ \Delta\alpha = \alpha_1  - \alpha_2 $ used in \cite{Geerits2023}) between the rays traveling along $ \widehat {ABD} $ and  $ \widehat {ACD} $. It is evident that following a phase shift of $2\pi$, the quantity $I_1(0) = I_1(2\pi)$, which is in complete accordance with the theory of dynamical diffraction in the context of neutron interferometry. Measurements of such interference patterns concerning
$ PS_1 $ rotation and observations of $ I_1 = I_1(\Delta\alpha) $ were not shown (or not performed)  in \cite{Geerits2023}.\\

\begin{figure}[hbtp]
\centering
\includegraphics[width=16cm]{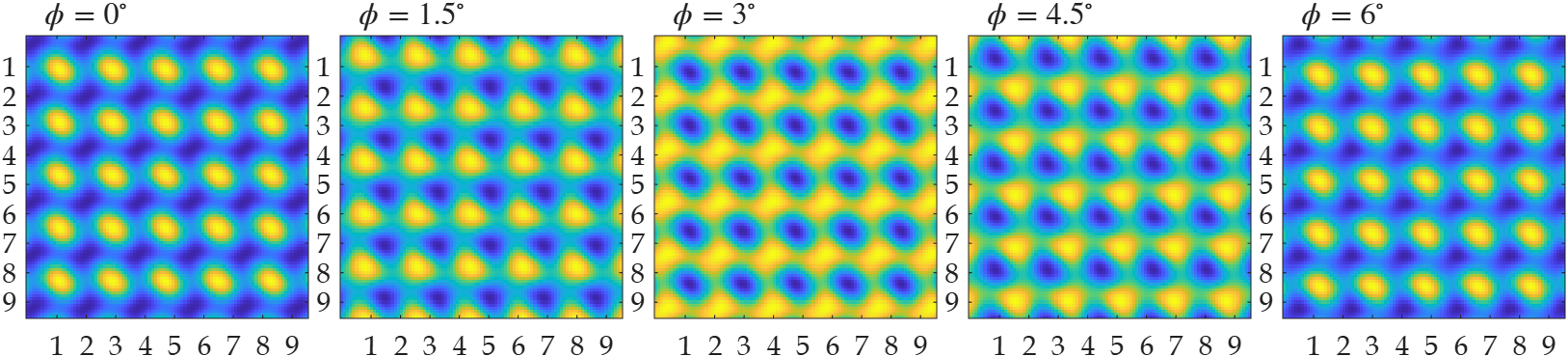}
\caption{$ PS_1 $  rotation $ \phi $: The rotation of the phase flag $ PS_1 $  causes the phase shift $ \Delta\alpha = \alpha_1 - \alpha_2 $ and a change in the interference pattern of the O beam $ I_1 $, {see} \cref{fig:4PlateIFM_II}.}
\label{fig:flag rotation}
\end{figure}

Finally, it must be emphasized that any attempt to detect an intrinsic or extrinsic neutron OAM using a crystal neutron interferometer is bound to fail, since the lateral coherence lengths $ \sigma_x, \sigma_z $  which must be identical for an n-OAM, differ here by about two orders of magnitude, and thus interference occurs only in the scattering plane. However, our simple derivation of the output {intensity calculations and} expectation value $\langle L_z \rangle   $  explains perfectly the measurement data in \cite{Geerits2023}. One must strongly question the interpretation given in Geerts \textit{et al.}.\\



\setstretch{1.18}

\section{Discussion}

A neutron interferometer always has only a single neutron, which is best described by a localized quantum state, as a wave packet. A normalized single-particle state satisfies the condition $ \int |\psi(\mathbf{r},t)|^2 \, d^3r = 1, $ 
where the wave function $\psi(\mathbf{r},t)$ contains complete information about the particle's spatial localization and momentum distribution.
To generate an n-OAM and thus vortices, once the   phase object must be coherently illuminated.
Cappelletti and Vinson   formulated this point explicitly in their analysis of the restriction of OAM by transverse coherence~\cite{Cappelletti2020}. They showed that the generation of single-particle vortex states with a vortex phase plate relies on each particle adequately sampling the area of the phase plate. If the transverse restriction of the incident wave prevents the wave packet from coherently sampling the central discontinuity of the vortex phase plate, the probability of obtaining a single-particle OAM state can vanish.  \\

The transverse coherence length must therefore be sufficiently large compared with the characteristic transverse scale of the phase structure that is responsible for the OAM transformation.  This does not mean that the entire neutron beam has to be globally coherent across its complete transverse diameter. Rather, the relevant requirement concerns the coherent transverse extent of the individual wave packet interacting with the phase object.

Given the very different lateral coherence lengths in neutron interferometry, one must consider only the portion of the generated vortex in which the lateral coherence lengths,  $\sigma_x $ and $\sigma_y$,  are of equal size, i.e., a maximum of 100~nm, which definitely cannot be measured using neutron interferometry and thus does not  explain the interference pattern measured. \\

Another reason to doubt that a neutron vortex lattice was observed in \cite{Geerits2023} is the following.
The generation of a vortex lattice requires conditions such as isolated zeros of the complex wave function, non-trivial phase circulation around these zeros and topologically protected phase winding  \cite{Bliokh2017,Soskin2001}. That means
(a)  the intensity is zero,
(b) the phase is undefined,
(c)  the phase increases by an integer multiple of $2\pi$ as it orbits the core.\\

One method of generating such states in laser optics is based on the interference of multiple coherent laser beams and using the Wollaston prism, as demonstrated in Kurzynowski (2006, 2010, 2012). This method was simulated using an LLLL neutron interferometer in Geerits (2023). 
From experimental layout the interference pattern caused by two wedges in a neutron interferometer cannot create phase singularities as with laser beams, because phase wedges primarily generate linear phase gradients rather than azimuthal phase structures, they naturally produce momentum shifts and interference fringes rather than intrinsic vortex states.

The two phase wedges in a LLLL neutron interferometer can indeed generate a highly structured interference pattern with periodic intensity modulation and spatially varying phase, but the generated structure corresponds primarily to an interference or Moir\'e pattern produced by controlled phase gradients, which is fundamentally different from a field containing a periodic array of phase singularities.
A periodic arrangement of phase singularities (vortex lattices) observed in an interference pattern would have to exhibit a structured wave field with local phase circulation, and phase singularities were not proven in \cite{Geerits2023}. 
An observation alone does not prove that every neutron is prepared in an intrinsic neither in an extrinsic orbital-angular-momentum eigenstate and  this should not automatically be interpreted as a vortex lattice in the topological sense.

Therefore, it seems obvious that the image in Fig. 4 of Geerits (2023) is simply the superposition of two phase patterns caused by two mutually perpendicular phase-shift wedges, as we calculated using the standard theory of dynamic neutron diffraction.\\

Similar arguments have already been published elsewhere \cite{Treimer2023,Treimer2024,Treimer2025}, showing that any experimental evidence for generating neutron orbital angular momentum (OAM) using a crystal neutron interferometer can be refuted. Each presented neutron interference pattern could always  be easily calculated using the established theory of dynamic neutron diffraction; therefore, it was unnecessary to interpret them as n-OAM.\\

\section{Conclusions}


The purpose of this paper is to provide a critical discussion and commentary on the interpretation of the article `Phase Vortex Lattices in Neutron Interferometry' by  {Geerits {\it{et al.}}}~\cite{Geerits2023}.
We provide a compact and consistent theoretical phase-based treatment of the four-plate perfect-crystal neutron interferometer.
This interferometer is described as two coupled LLL loops, resulting in a forward intensity $I_1$ (see Fig.~2) that is determined by three coherent interference terms.
Phase objects such as wedges, when inserted into one or two of the LLL arms, produce structured fringe patterns whose qualitative behavior depends heavily on whether one or two wedges are present and where they are placed.\\
The derived intensity expressions reproduce known results from Heinrich \textit{et al.} and Geerits \textit{et al.}, and they agree with our calculations which did not take into account an OAM. 
The interference patterns we calculated result from the simple superposition of coherent partial waves, which ultimately leads to an output intensity as shown in Fig. 4 in \cite{Geerits2023}. \\
The interpretation of the observed interference pattern as a neutron vortex lattice and, consequently, as evidence for extrinsic neutron OAM relies on additional assumptions that are not uniquely implied by the experimental data.

\section{Acknowledgements}

 The work of author TJ was funded by the National Institute of Standards and Technology,
 U. S. Department of Commerce.

\end{document}